\documentclass[aps,prb,twocolumn,superscriptaddress,showpacs]{revtex4-1} 

\usepackage{amssymb}
\usepackage{graphicx, color}
\usepackage{amsmath}
\usepackage{dsfont}
\usepackage{bbm}
\usepackage{natbib}
\begin{document}

\title{Effects of finite superconducting coherence lengths and of phase gradients in topological SN and SNS junctions and rings}

\author{D. Chevallier}
\affiliation{Laboratoire de Physique des Solides, CNRS UMR-8502, Universit\'e Paris Sud, 91405 Orsay Cedex, France}
\author{D. Sticlet}
\affiliation{Laboratoire de Physique des Solides, CNRS UMR-8502, Universit\'e Paris Sud, 91405 Orsay Cedex, France}
\author{P. Simon}
\affiliation{Laboratoire de Physique des Solides, CNRS UMR-8502, Universit\'e Paris Sud, 91405 Orsay Cedex, France}
\author{C. Bena}
\email{cristinabena@gmail.com}
\affiliation{Laboratoire de Physique des Solides, CNRS UMR-8502, Universit\'e Paris Sud, 91405 Orsay Cedex, France}
\affiliation{Institute de Physique Th\'eorique, CEA/Saclay, Orme des Merisiers, 91190 Gif-sur-Yvette Cedex, France}

\date{\today}

\begin{abstract}
We study the effect of a finite proximity superconducting (SC) coherence length in SN and SNS junctions consisting of a semiconducting topological insulating wire whose ends are connected to either one or two s-wave superconductors. We find that such systems behave exactly as  SN and SNS junctions made from a single wire for which some regions are sitting on top of superconductors,  the size of the topological SC region being determined by the SC coherence length. Moreover, we study the effects of continuous phase gradients in both an open and closed (ring) SNS junction. We find that such phase gradients  play an important role in the spatial localization of the Majorana fermions.
\end{abstract}


\maketitle

\section{Introduction} \label{sec:introduction}
  
There has been a lot of recent interest in the possibility to realize and detect Majorana fermionic states (see Refs. \onlinecite{beenakker} and \onlinecite{alicea} for recent reviews).
Following earlier theoretical proposals,\cite{lutchyn_dassarma,oreg_vonoppen} semiconducting wires with a strong spin orbit coupling, such as InAs and InSb, in proximity of an s-wave superconductor have become a promising platform to observe Majorana fermions. The properties of such systems
have been extensively studied for homogeneous superconductors \cite{stoudenmire,simon_loss,cook_franz,stanescu_dassarma,bena_sticlet,lutchyn_fisher} but also in
superconductor-normal (SN) junctions \cite{wimmer,gibertini,cssb_sns,prada,sau_dassarma,klinovaja,pientka,rainis} and
superconductor-normal-superconductor (SNS) junctions.\cite{kitaev,kwon,fu_kane,tanaka,ioselevich,law_lee,liang,badiane,black-schaffer,vanheck11,sanjose}

 A few recent experiments have reported the observation of zero-bias conductance peaks,\cite{kouwenhoven,xu,heiblum}
the fractional ac Josephson effect in the form of doubled Shapiro steps, \cite{rokhinson} both consistent with the existence of Majorana fermions. 
However such experimental observations have been shown to also have simpler explanations: a non-quantized zero bias peak could also arise in principle without any Majorana fermions
 \cite{lee,aguado_kondo,bagrets} while  the fractional ac Josephson effect can also occur in conventional SNS junctions  and be understood simply in terms of Landau-Zener processes associated with the Andreev-bound-state spectrum of the junction.\cite{sau} Therefore a  non-equivocal proof for the observation of Majorana fermions is still lacking. 

Along the lines of finding the most appropriate experimental strategy to detect such states, in a previous paper \cite{cssb_sns} we have analyzed the possibility to detect the Majorana states in NS junctions and SNS junctions using local energy-resolved spectroscopy such as STM.  We focus here on analyzing the effects of some relevant experimental factors such as a non-zero penetration length, as well as a non-zero uniform phase gradient for the detection of such states in linear as well as ring Josephson junctions.

In a previous work \cite{cssb_sns} we have shown  that the formation of Majorana modes in a finite-size normal link can be controlled by varying the phase difference between two superconductors (SCs) in an SNS junction. Such Majorana modes stem from the intrinsic Andreev bound states residing in such link and thus may extend over the entire normal link. Such numerical prediction has been subsequently confirmed by Klinovaja and Loss \cite{klinovaja} who obtained a full analytical expression  of the Majorana wave function in an SN junction.  These analyses have focused on sharp NS interfaces, for which the SC order parameter varies discontinuously at the interface. In the present work we consider first a non-zero penetration length of the SC in the normal state, and therefore a continuous decay of the SC order parameter close to the interface. We use a tight-binding exact-diagonalization technique to describe a semiconducting wire with strong spin-orbit coupling, end-contacted to s-wave SCs, in the presence of a Zeeman field \cite{alicea, cssb_sns}. We find that for small penetration lengths with respect to the length of the wire, a normal region is preserved, and the Majorana modes arising in this region have a significant spatial extension. However, for larger penetration lengths, the entire wire becomes a topological SC, and localized Majorana fermions form only at its ends. 

We also study the effect of a uniform gradient of the SC phase in both a linear and ring Josephson junction (see Fig. \ref{fig:setup}). We find that, consistent with previous observations on infinite wires, for small phase gradients the system remains fully topological, while for larger phase gradients, the portion of the system subject to the gradient enters into a gapless phase which does not support extra Majorana states besides the end Majorana states.  
For a Josephson ring we find that Majorana states can be induced in the normal region if the phase gradient is not too small, and if the total phase twist is close to an odd multiple of $\pi$. The periodicity of the energy modes with the total phase twist has a small shift from $2\pi$, shift that goes to zero for large systems. Similar to the linear case, the system enters into a gapless phase when the phase gradient becomes larger than a critical value.

The paper is organized as follows. In Sec. II we present the considered setups and the corresponding tight-binding models. In Sec. III we study the effect of the SC coherence length on the location and spatial extension of the Majorana fermions in the wire. Finally, in Sec. IV, we focus on the effects of a phase gradient for linear and ring SNS junctions. We conclude in Sec. V.

\section{Model} \label{sec:model}
In this section we present two models: one corresponding to linear SN and SNS junctions which will be used to understand the effect of the finite  superconducting coherence length (Part A), and the second one corresponding to a SNS ring junction which will allow us to study the effect of a phase gradient (Part B).

\subsection{Linear NS and SNS junctions} \label{sec:linear}

\begin{figure}[ht]
	\centering
		\includegraphics[width=6.5cm]{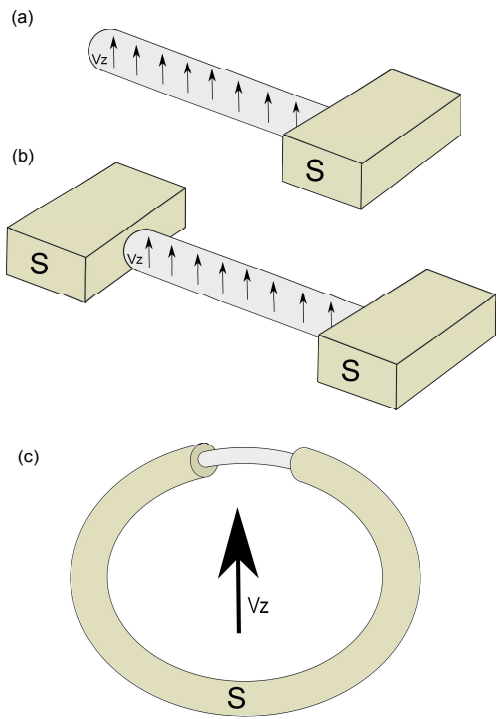}
	\caption{(Color online) Scheme of the proposed setups : a) a semiconducting nanowire connected to an s-wave SC b)  a semiconducting nanowire connected at both ends to s-wave SCs c) a semiconducting nanowire connected to an s-wave SC ring.}
	\label{fig:setup}
\end{figure}

\begin{figure}[ht]
	\centering
		\includegraphics[width=6.5cm,height=6cm]{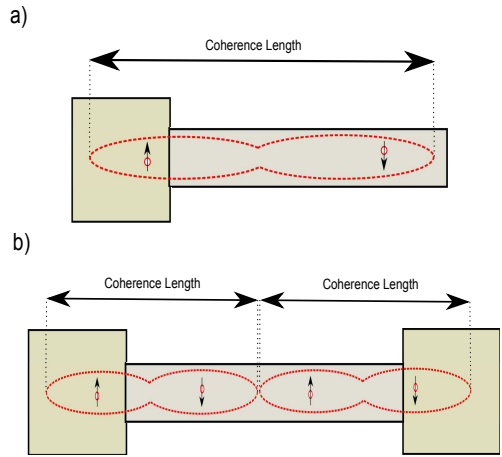}
	\caption{(Color online) The coherence length for a superconductor (green) corresponding to the "size" of a Cooper pair. This coherence length acts as a SC penetration length in the normal part (gray).}
	\label{fig:coherence}
\end{figure}

\begin{figure}[ht]
	\centering
		\includegraphics[width=7.5cm,height=5.5cm]{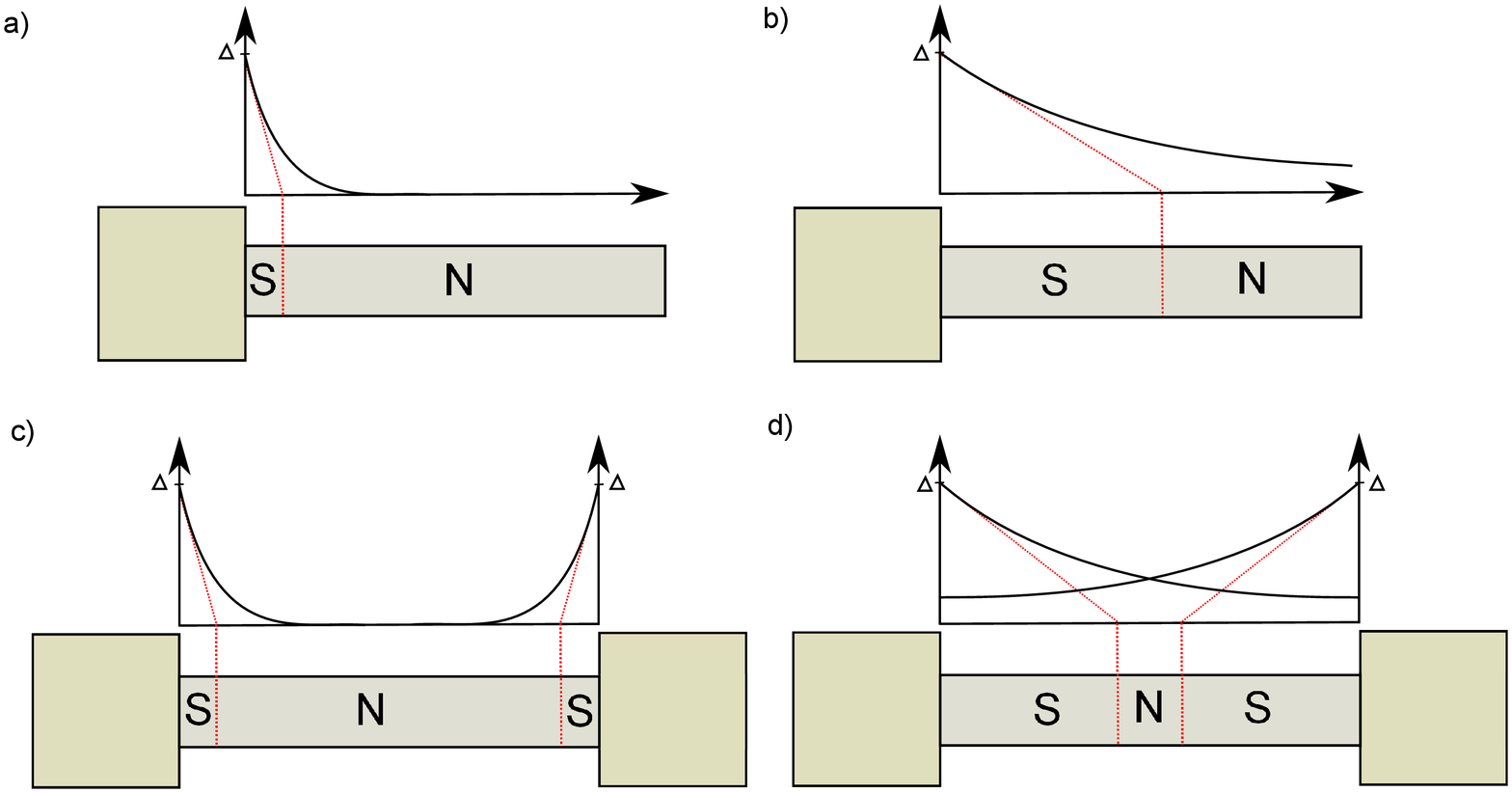}
	\caption{(Color online) A schematic depiction of the spatial dependence of the SC order parameter in NS and SNS junctions. The red dotted lines indicate the `soft' effective NS boundaries at a distance $\xi_{sc}$ from the interface. }
	\label{cl}
\end{figure}

Our starting point is a semiconducting nanowire with strong Rashba spin-orbit (SO) coupling.\cite{oreg_vonoppen,stoudenmire}  In the presence of a Zeeman magnetic field and in the proximity of a SC, such a wire has been shown to exhibit end Majorana fermionic states. \cite{kitaev,lutchyn_dassarma,oreg_vonoppen}  We focus on using such a wire to make SN or SNS long junctions by end-contacting it to one or two s-wave SCs (See Fig.~\ref{fig:setup} a) and b)).
In the semiconducting wire the superconductivity is induced only for distances from the NS interface smaller than the SC coherence length; here the coherence length acts as a superconducting penetration length, see Figs.~\ref{fig:coherence} and \ref{cl} for an intuitive representation of this coherence length, as well as for a schematic depiction of the spatial dependence of the SC order parameter in the studied junctions. 

To be able to perform a numerical analysis of this system, we model it as a 1D tight-binding chain \cite{stoudenmire,likharev}
\begin{eqnarray}\label{hamiltonian}
{\cal{H}}&=&\sum^N_{j=1}c^\dagger_{j}((t-\mu)\tau_z+V_z\sigma_z) c_{j}-\frac{1}{2}c^\dagger_{j}(t+i\alpha\sigma_y)\tau_z c_{j+1}+h.c.\notag\\
&&+\sum_{j=1}^{N}\Delta_1 (1-\frac{j}{N}) e^{-j/\xi_{sc}} e^{i \phi_1} c_{j\uparrow}c_{j\downarrow}+h.c.\notag\\
&&+\sum_{j=1}^N\Delta_2 (\frac{j}{N})e^{(j-N)/\xi_{sc}} e^{i\phi_2} c_{j\uparrow}c_{j\downarrow}+h.c.\notag\\
\end{eqnarray} 
where the sum is performed only over the $N$ sites belonging to the semiconducting wire (we do not include the two s-wave SC sites in the model, but we assume that the sole effect of the SCs is to induce a SC order parameter in the wire which decays exponentially with the distance from the NS interface). Also, $\sigma$ and $\tau$ are Pauli matrices which act respectively in the spin and particle-hole spaces, $\mu$ is the chemical potential, $V_z$ the Zeeman field, $\alpha$ the SO coupling, $\Delta_{1/2}$ are the SC gaps of the two SCs, and $\xi_{sc}$ is the coherence length of the superconductors (we assume that the two SCs have the same coherence length). According to K. K. Likharev \cite{likharev}, the factors $(1-\frac{j}{N})$ and $\frac{j}{N}$ before the superconducting terms are introduced in order to smooth out the transition of the effective superconducting gap between the two superconductors in an SNS junction (these factors are obviously not present in an NS junction).  We consider a perfect transmission of the junction(s) throughout all the calculations.
For the NS junction depicted in Fig. \ref{fig:setup} a) we have taken $\Delta_2=0$ and for the SNS junction depicted in Fig. \ref{fig:setup} b) we have taken $\Delta_{1}=\Delta_2=\Delta$. We work in units of $t=1$, and the superconducting penetration length $\xi_{sc}$ is given in units of the lattice constant $a$ which we also take to be equal to $1$.

\subsection{SNS ring junction}

We consider using a topogical semiconducting wire to make an SNS ring junction (See Fig. \ref{fig:setup} c)).
The superconductivity is induced via the proximity effect in one part of the ring. This ring SNS junction can also be modeled using a tight-binding chain
\begin{eqnarray}\label{ring_hamiltonian}
{\cal{H}}&=&\sum^{\hookleftarrow}_{j}c^\dagger_{j}((t-\mu)\tau_z +V_z\sigma_z)c_{j}-\frac{1}{2}c^\dagger_{j}(t+i\alpha\sigma_y)\tau_z c_{j+1}+h.c.\notag\\
&&+\sum_{j=N_1}^{N_2}\Delta e^{i(N_1-j) \nabla\phi} c_{j\uparrow}c_{j\downarrow}+h.c\notag\\
\end{eqnarray} 
where $\hookleftarrow$ corresponds to summing over all the sites around the ring. The topological SC section lies between sites $N_1$ and $N_2$. All the other sites correspond to the non-SC semiconducting nanowire. The phase gradient is modeled as a $\nabla\phi$ phase variation between two neighboring sites $j$ and $j+1$ in the SC region. All the other parameters are identical to the ones presented in the previous section.

\subsection{LDOS and Majorana polarization reminder}

The quantities that we use to describe the above setups are the local density of states (LDOS) and the local $x$-axis Majorana polarization which are given by the following expressions~\cite{bena_sticlet}
\begin{align}
\rho_j(\omega)&=\sum^{4N}_{i=1}\sum_{\delta=\uparrow,\downarrow}\delta(\omega-E_i)\left|\alpha^i_{j,\delta}\right|^2\\
{\cal{P}}^{x}_j(\omega)&=\sum^{4N}_{i=1}\sum_{\delta=\uparrow,\downarrow}\delta(\omega-E_i)2\textrm{Re}(\alpha^{i*}_{j,\delta}\beta^{i}_{j,\delta}),
\end{align}
where $(\alpha^i_{j\uparrow},\beta^{i}_{j\uparrow},\alpha^{i}_{j\downarrow},\beta^{i}_{j\downarrow})$ are the components of the wave function on site $j$ for the $i^{th}$ eigenstate of the system in the $(c^\dagger_{j\uparrow},c_{j\uparrow},c^\dagger_{j\downarrow},c_{j\downarrow})$ basis. An exact digonalization of the Hamiltonians (\ref{hamiltonian}) and (\ref{ring_hamiltonian}) allows us to evaluate these two quantities. We focus on the limit when the system is in the topological phase by choosing $V_z=0.4, \Delta=0.3, \mu=0, \textrm{and}\; \alpha=0.2$.
A finite width $0.0002 \hbar v_F/a$ for the delta functions is introduced in the numerical evaluations.

\section{Effect of the SC coherence length on the localization of Majorana fermions}    
    
In this section, we study NS and SNS junctions made of a semiconducting nanowire connected to either one or two superconductors. The superconductivity in the nanowire is induced solely via the proximity effect (see Fig. \ref{fig:coherence} and \ref{cl}), with the SC order parameter decreasing exponentially with the distance from the SC interface ($\propto e^{-\frac{x}{\xi_{sc}}}$).

\begin{figure}[ht]
	\centering
		\includegraphics[width=8cm]{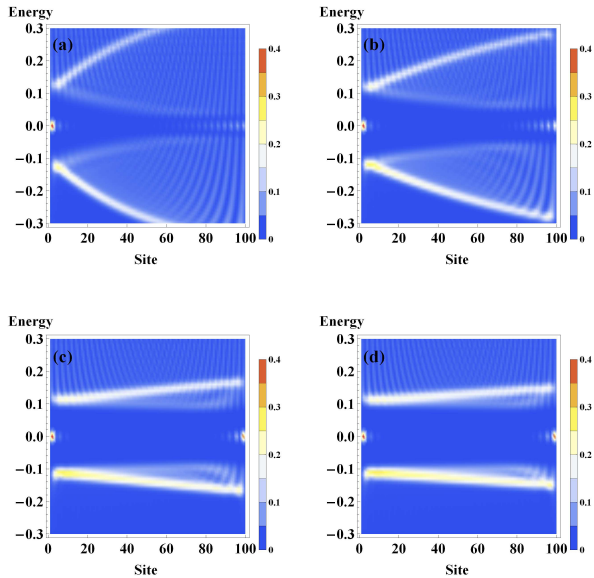}
	\caption{(Color online) LDOS for an NS junction as a function of energy and position for various values of $\xi_{sc}$. From left to right and up to down: $\xi_{sc}$=50, 100, 400, 600.}
	\label{fig:LDOS1}
\end{figure}

\begin{figure}[ht]
	\centering
		\includegraphics[width=8cm]{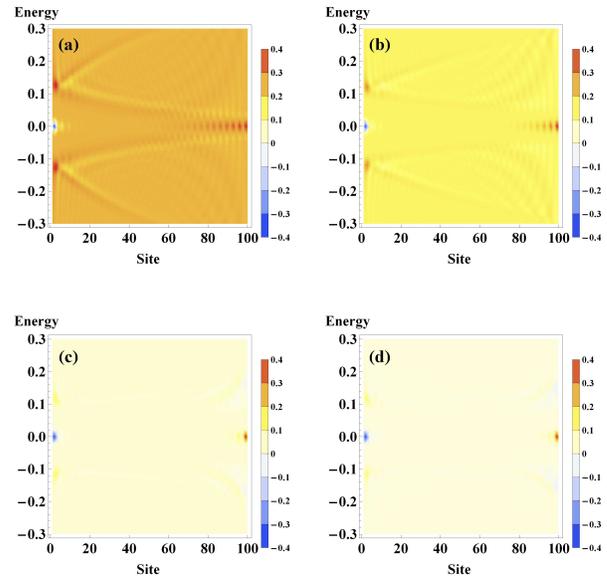}
	\caption{(Color online) Majorana polarization for an NS junction as a function of energy and position for various values of $\xi_{sc}$. From left to right and up to down: $\xi_{sc}$= 50, 100, 400, 600.}
	\label{fig:majopola1}
\end{figure}

\emph{NS junction} We first study an NS junction (see Fig. \ref{fig:setup} (a)). In Fig. \ref{fig:LDOS1} and \ref{fig:majopola1}, we plot the LDOS and the Majorana polarization of the nanowire as a function of energy and position for various values of the penetration length.

In Fig. \ref{fig:LDOS1}, we observe the formation of a zero energy mode close to the left end of the nanowire, in the vicinity of the NS interface, for values of $\xi_{sc}$ which are not too small (for very small values of $\xi_{sc}$, for example $\xi_{sc}=10$, precursor Majorana states form at the end of  the wire, but these states are not fully Majorana polarized, and their energy is small but non-zero, and varies with the different parameters in the system). Also, Majorana states form at the right end of the wire, but their spatial extension depends on the value of the superconducting penetration length. Thus, for small values of the superconducting penetration length, the right-hand-side zero mode is extended over a large region of the wire (see the first two panels in Fig.~\ref{fig:LDOS1}). With increasing the value of the superconducting penetration length, this zero-energy state becomes more and more localized towards the end of the wire. 

To confirm that these states correspond to actual Majorana fermionic states we plot the Majorana polarization along the x-axis. 
Indeed, Fig.~\ref{fig:majopola1} shows that the left-end zero-energy state is a Majorana fermion with a total polarization of $-1$ (same as in Ref.~\onlinecite{bena_sticlet,cssb_sns} we obtain the total Majorana polarization by integrating over the entire peak) while the right-end extended Majorana carries a $+1$ polarization, consistent with the conservation of the total Majorana polarization in the nanowire.  

The spatial extension of the Majorana states can be understood in relation with the analysis of the NS junction in Ref.~\onlinecite{cssb_sns,klinovaja}, and by assuming that the topological SC region induced in the wire is roughly given by the size of the SC coherence length. Moreover one can consider that an effective `soft' NS boundary is created at a distance from the NS interface equal to the SC coherence length (see Fig.~\ref{cl}). Thus, when the penetration length is much smaller than the nanowire length, the NS boundary is close to the left end of the wire, and the right Majorana mode extends over the large entire normal region. When the coherence length increases, the effective SC region becomes larger, the soft NS boundary is moving towards the right end of the wire, and the right-hand Majorana mode becomes more and more localized. When $\xi_{sc}$ is larger than the length of the wire $l$, the entire wire is in a topological SC state, the Majorana states are localized at the two ends of the nanowire, and the extension of the Majorana wavefunction is constant and independent of the length of the nanowire ($\xi_{sc}>l>\xi_{m}$ where $\xi_{m}$ is the "size" of the Majorana wavefunction).

\begin{figure}[ht]
	\centering
		\includegraphics[width=7cm]{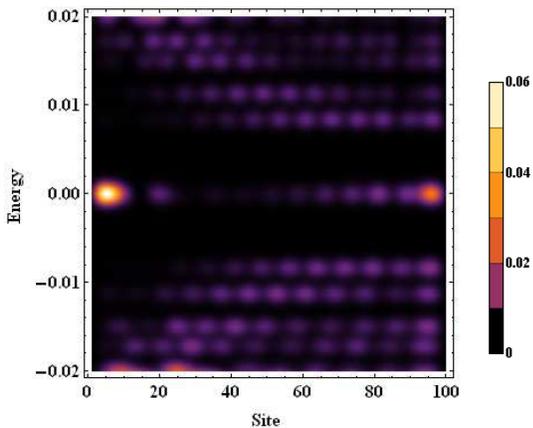}
	\caption{LDOS as a function of energy and position for $V_z=0.03$, $\Delta=0.02$, $\alpha=0.2$ and $\xi_{sc}\approx1/\Delta\approx50$ : two uncoupled Majoranas are protected by a topological gap $\Delta_T =V_z -\Delta\approx0.01$.}
	\label{realistic}
\end{figure}

Up to now and throughout most of the paper we consider for simplicity the superconducting penetration length (i.e. coherence length) as independent of the superconducting gap. However, this penetration length is related to the superconducting gap, $\xi_{sc}=\hbar v_F/\Delta$. Taking into account the relation between the penetration length and the gap would affect the physics described here only quantitatively, but not qualitatively: an increase in $\xi_{sc}$ would correspond to a decrease in the SC gap, but one can remain in a qualitatively similar topological phase with a renormalized effective topological gap by adjusting the other parameters in the model such as the Zeeman magnetic field. To confirm this, in Fig. \ref{realistic} we have plotted the LDOS as a function of energy and position for a set of parameters where $\Delta\approx1/\xi_{sc}$. We can see two uncoupled Majorana states protected by a topological gap given by $\Delta_T =V_z -\Delta$. The first Majorana state is completely localized at the left end of the nanowire and the second one is extended over the entire normal part. We can thus claim that the qualitative features of the Majorana physics described using the assumption that the penetration length is a free parameter will not be affected by this assumption.

\begin{figure}[ht]
	\centering
		\includegraphics[width=8cm]{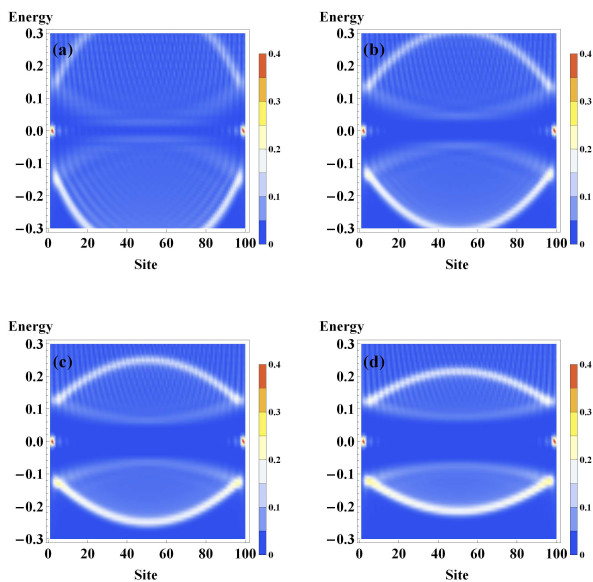}
	\caption{(Color online) LDOS for an SNS junction as a function of energy and position for $\phi=0$ and various values of $\xi_{sc}$. From left to right, and up to down: $\xi_{sc}$=20, 40, 70, 100.}
	\label{fig:LDOS2}
\end{figure}

\emph{SNS junction} We now turn to the analysis of SNS junctions (see Fig.~\ref{fig:setup} b)). In order to be as clear as possible, we present a plot of the LDOS for an SNS junctions with a zero phase difference, and a plot of the Majorana polarization for an SNS junction with a phase difference of $\pi$. Thus, for a zero phase difference, in Fig.~\ref{fig:LDOS2}, we show the LDOS as a function of energy and position for various values of the superconducting penetration length (modeled as before by an SC order parameter exponentially decreasing ($\propto e^{-\frac{x}{\xi_{sc}}}$) with the distance from the two NS interfaces).
 
We can see that, similar to the NS junction, effective NS `soft' boundaries form inside the normal region 
for small values of the superconducting penetration length ($\xi_{sc}<l/2$). In this regime, the situation is equivalent to the `hard' SNS junction described in  Ref.~\onlinecite{cssb_sns}. Indeed we note the formation of two zero-energy states at the ends of the wire, plus the formation of finite-energy quasi-Andreev bound states in the effectively-normal region.  The two Majorana states at the ends of the wire have opposite polarization ($\pm1$), with no Majorana polarization in the bulk of the wire (not shown). When we increase the superconducting penetration length, these Majorana states become more and more localized towards the two ends of the nanowire and their spatial extension is constant and independent of the nanowire length ($\xi_{sc} > l/2 >\xi_{m}$). The normal region is becoming smaller and smaller, and the critical value of the superconducting penetration length for which the entire wire becomes SC is given by half of the nanowire length (see Figs.~\ref{fig:coherence}, and \ref{cl}). Above this value no normal region forms inside the wire.

\begin{figure}[ht]
	\centering
		\includegraphics[width=8cm]{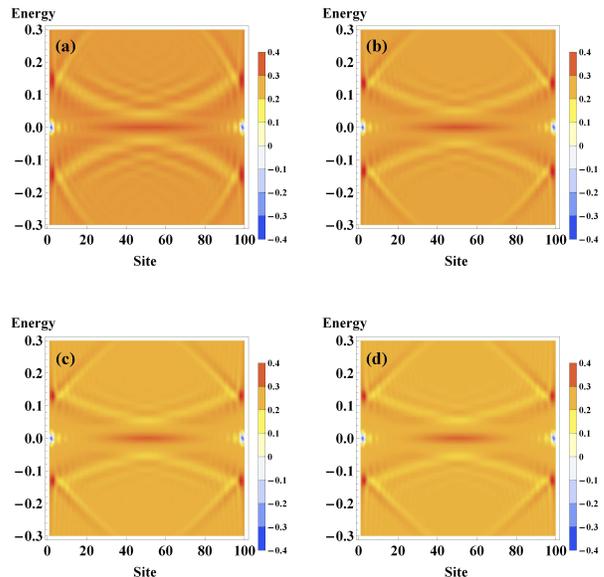}
	\caption{(Color online) Majorana polarization for a SNS junction as a function of energy and position for a phase difference of $\phi=\pi$ and various values of $\xi_{sc}$. From left to right, and up to down: $\xi_{sc}$=20, 40, 70, 100.}
	\label{mpsns}
\end{figure}

The situation is more complex when the phase difference is non-zero. As described in Ref.~\onlinecite{cssb_sns}, in a long SNS junction with a non-zero phase difference, quasi-Majorana modes are being form inside the normal region, and these modes reach a zero-energy and have a full Majorana character when the phase difference is equal to $\pi$. Here we expect this behavior to hold for small values of the coherence length $\xi_{sc}<l/2$, and indeed, as plotted in Fig.~\ref{mpsns}, and consistent with Ref.~\onlinecite{cssb_sns},  for $\phi=\pi$ we observe the formation of the zero-energy modes in the center of the wire; we have checked that these modes have a total polarization of $-2$, opposite to the two `parallel' Majorana modes at the two ends of the wire which have a polarization of $+1$ each.

For larger superconducting penetration lengths the central normal region becomes smaller and smaller, and for superconducting penetration lengths larger than the critical value ($\xi_{sc}>l/2$) no purely-normal region exists in the middle of the wire. However, due to the phase difference of $\phi=\pi$ the effective order parameter penetrating from the two ends of the wire will always cancel in the center of the wire \cite{likharev} (we assume that the two SC order parameters and superconducting penetration lengths are identical for the two SCs, if it is not the case, the point where the effective SC order parameter will cancel will be situated at a different position in the wire). Thus one will always form a non-topological region inside the wire in the region where the effective value of the SC order parameter is small, and consequently two Majorana states inside this region with polarization opposite to that of those at the ends of the wire.

The coherence length, which is given roughly by the extension of a Cooper pair, can greatly vary, from a few microns in an usual superconductor (Al) to a few nanometers in the case of cuprates. While nanowires a few hundred nanometers in length are easy to obtain (typically the minimum size is around $\propto 500$nm), we can expect that the transition $\xi_{sc}>l/2$ may be achieved in typical setups. Moreover, the setup we propose here in which the proximity effect is achieved via the end contact with a superconductor, rather by placing the nanowire on top of a superconductor, may be easier to realize, and thus have technological advantages in the realization of topological SNS junctions.

\section{The Josephson effect in the presence of a uniform phase gradient}

In a recent paper,\cite{Romito} it was shown that supercurrents in the bulk of the superconductor could be used in principle to manipulate the Majorana fermions. More precisely, it has been shown that a constant spatial gradient in the phase of the superconducting parameter can drive the system from a topological phase supporting Majorana fermions to a trivial phase without zero-energy bound modes. Here we use the tight-binding model to investigate the Josephson effect in linear and ring geometries in the presence of a phase gradient.

\begin{figure}
\includegraphics[width=0.8\columnwidth]{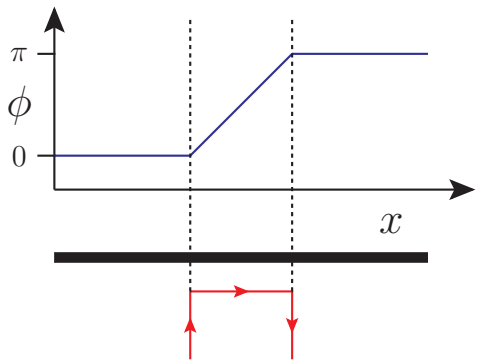}
\caption{(Color online) Variation of the superconducting phase in a topological SC wire:  the phase $\phi$ is set to zero in the left region, and a constant gradient is considered to be induced in the central region by bulk supercurrents such that the phase $\phi$ reaches a value of $\pi$ in the left region of the wire.}
\label{fig:sup}
\end{figure}

For the linear geometry, we consider a system similar to the long SNS junction explored in Sec. \ref{sec:linear}. However, in this section we consider that the superconducting gap, the spin-orbit coupling as well as the Zeeman field are uniform over the entire wire, the only parameter depending on position being the superconducting phase. A typical phase spatial dependence is presented in Fig.~\ref{fig:sup}, such that there are two outer regions of constant phase, with a central region uniformly experiencing a phase twist. We are interested to find if the uniform phase variation may give rise to a physics similar to the one observed in a traditional SNS junction. \cite{cssb_sns} We also consider a semiconducting ring with a ``normal'' region, and the rest experiencing a superconducting proximity effect and an uniform phase gradient. 

\subsection{Constant phase gradient in a wire}
Before launching into the numerical analysis for the tight-binding form of the Hamiltonian in Eq.~(\ref{ring_hamiltonian}), we present an analytical study of the effect of the constant phase gradient on the topological index, that allows one to predict the existence of a topological phase transition. It is not entirely surprising that the gradient of the superconducting parameter can make the system switch between a topological trivial and nontrivial phase. For a uniform gradient, this can be readily understood in the limit of an infinite wire. As shown in Ref. \onlinecite{Romito}, the phase of the pairing term can be gauged away, with the effect of adding a gradient-dependent correction to the canonical momentum and a renormalization of the hopping parameter, as well as of the spin-orbit coupling. For the infinite wire, the condition to have a topological phase is given by\cite{Romito}
\begin{eqnarray}\label{topcond}
\bigg\{\bigg[\mu-\frac{t}{8}(\nabla\phi)^2\bigg]^2-\frac{\alpha^2(\nabla\phi)^2}{4}-V_z^2
+|\Delta|^2\bigg\}&&\notag\\
\times\bigg\{\bigg[\mu-2t-\frac{t}{8}(\nabla\phi)^2\bigg]^2-\frac{\alpha^2(\nabla\phi)^2}{4}-V_z^2
+|\Delta|^2\bigg\}
&<&0.\notag\\
\end{eqnarray}
When the bandwidth $t$ is larger than the other parameters of the system, the second term of the product is always positive. Thus, for a zero chemical potential $\mu$, the critical phase gradient is the exact lattice analogue of the continuum expression determined in Ref.~\onlinecite{Romito}
\begin{equation}\label{critphi}
(\nabla\phi)_c=2\sqrt{2}
\bigg\{\bigg(\frac{\alpha}{t}\bigg)^{\!2}
+\bigg[\frac{V^2_z-\Delta^2}{t^2}
+\bigg(\frac{\alpha}{t}\bigg)^{\!4}
\bigg]^{1/2}
\bigg\}^{1/2}.
\end{equation}
A phase gradient has a Cooper pair breaking effect and can close the superconducting gap at the Fermi momentum. This leads to a second critical value for the phase gradient $(\nabla\phi)_{gl}$ above which the bulk gap closes, and the system enters a gapless regime. Its exact value is determined by numerically studying the closing of the gap for an infinite system that experiences a uniform phase gradient $\nabla\phi$. The momentum space Bogoliubov-de Gennes Hamiltonian is given by
\begin{eqnarray}
H&=&\frac{1}{2}\sum_k C_k^\dag\mathcal HC_k,\quad
C_k=(c^\dag_{k\uparrow},c^\dag_{k\downarrow},
c_{-k\downarrow},c_{-k\uparrow})
\notag\\
\mathcal H&=&(t-\mu)\tau_z-[t\cos(k)
+\alpha\sin(k)\sigma_y]\cos(\nabla\phi/2))\tau_z\notag\\
&&-[t\sin(k)-\alpha\cos(k)\sigma_y]\sin(\nabla\phi/2)+V_z\sigma_z-|\Delta|\tau_x\notag\\
\end{eqnarray}
For our set of system parameters $(V_z=0.4,\Delta=0.3,\alpha=0.2,\mu=0)$, the critical gradient for which the system enters into a gapless phase is $(\nabla\phi)_{gl}\simeq 0.27$, while the topological condition yields a critical phase gradient of $(\nabla\phi)_c\simeq 1.57$. Thus, for the parameters we consider, the system enters first into a gapless metallic region before reaching a gapped trivial phase.

We want to use this type of arguments to study a non-uniform wire with a variable-length central region that experiences the twist of the superconducting phase (we denote this region as gradient region - GR). When the phase gradient is zero, the system is in a topological phase with Majorana modes at the ends. The question that we are trying to address is if the central GR can be driven into a trivial gapped phase such that Majorana fermions can form also at the interface with the topological phases.

\begin{figure}[ht]
\includegraphics[width=\columnwidth]{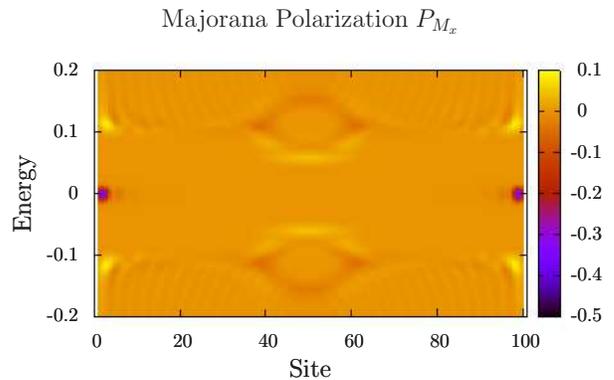}
\caption{(Color online) The phase of the superconducting parameter is twisted by $\pi$ over the $20$ central sites. The central regions remains topologically nontrivial and only two Majorana fermions with the same polarization form at the ends of the wire.}
\label{fig:mpolx20site}
\end{figure}

We perform  a numerical analysis for a system of 100 sites. We start with a small value for the phase gradient.
In Fig. \ref{fig:mpolx20site} the phase gradient over the 20 central sites is $\pi/20$, smaller than $(\nabla\phi)_{gl}$ value, and the GR remains topological. Hence there are only two Majorana fermionic states which form at the ends of the wire. Because of the relative phase $\phi=\pi$ between the two outer regions, the Majorana fermions have parallel Majorana polarizations.  We should note that the formation of two identical Majorana modes at the two ends of the wire does not raise a problem from the perspective of the conservation of the Majorana polarization, since, quite puzzling, the zero-energy Majorana polarization is compensated by a continuous distribution of Majorana polarization in the higher-energy bands (if we integrate the Majorana polarization over energy at any given site, the integral yields zero).

Let us start with a wire in the topological phase. The goal is to create new Majoranas by driving the central GR into a trivial gapped phase using only an increasing uniform phase gradient. We find numerically that this is not possible, and that the system is always driven first into a gapless phase $(\nabla\phi)_{gl}<(\nabla\phi)_c$. This result is consistent with the phase diagrams in Ref.~\onlinecite{Romito}. Therefore no Majorana fermions form at the interface between the GR and the exterior topological regions. 

\begin{figure}[ht]
\includegraphics[width=\columnwidth]{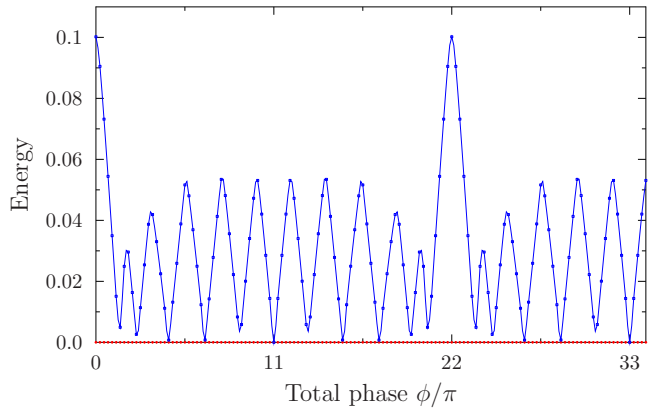}
\caption{(Color online) Evolution of the first two positive eigenvalues with the phase difference $\phi$ for a $11$-site GR.  The constant zero-energy mode corresponds to an end Majorana fermionic state. The energy of the other mode evolves with  $\phi$ and reaches zero only when there is a $(2n+1)\pi$ phase difference between two neighboring sites (we can see that this happens here for $n=0$ and $n=1$, corresponding to $\phi=11 \pi$ and respectively $\phi=33 \pi$).}
\label{eigwire}
\end{figure}

A special situation arises however when the gradient region is constructed as a series of Josephson junctions with a phase increase of $(2n+1)\pi$ between two neighboring sites, with $n$ being an integer (see Fig. \ref{eigwire}). The GR must consist of an odd number of sites, to ensure a relative phase difference $(2n'+1)\pi$ between the left and right regions. 
Such a situation can for instance be encountered in arrays of topological superconducting islands which has been shown in Ref.~\onlinecite{vanheck} to provide a macroscopic realization of the Kitaev model \cite{kitaev}.

\begin{figure}[ht]
\centering
\includegraphics[width=\columnwidth]{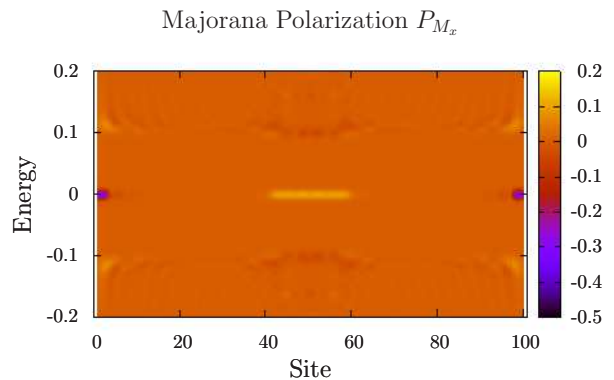}
\caption{(Color online) Phase gain of $11\pi$ over $11$ sites. Two Majorana fermions form in the GR having an opposite polarization with respect to the end modes.}
\label{fig:mpolx11site}
\end{figure}

In this situation, extended Majorana fermions can form in the GR. In Fig. \ref{fig:mpolx11site} we represent such a situation with two extended Majorana fermions forming for a phase difference of $\phi=11\pi$ over an $11$ site GR. Integrating the Majorana polarization over the central region yields a total value of two, showing that only two Majorana states form in this region. 

Note that the system is invariant under a change of $2\pi$ in the phase gradient. Hence for a gradient region of  $N$ sites, there is a $2\pi N$ periodicity in the total relative phase $\phi$ between the left and right ends of the wire. In the special case when the phase gradient is distributed over an odd number of sites with Majorana fermions forming at $\nabla\phi=(2n+1)\pi$, the periodicity in the total phase is $4\pi N$. 

To summarize this part, we have shown that an extended Majorana fermion forms in the gradient region when the nanowire is made of a series of Josephson junctions with a phase difference equal to an odd multiple of $\pi$ between each neighboring sites.

\subsection{Ring with a constant phase gradient}
Here we investigate the presence of Majorana fermions in a ring geometry divided in two regions of variable length. One of the regions is in a normal state (does not experience a SC effect), and the second one is SC with a SC phase that may be twisted. We investigate under what conditions Majorana fermions do form in the normal region.

In the absence of a phase gradient, there no Majorana fermions form inside the normal region. Even though the SC region is in a topological phase, the would-be Majorana modes at the two ends communicate through the normal region. Therefore they split and become normal fermionic modes lying in the SC gap. We have checked that for a non-zero but imperfect coupling between the normal and the SC region, quasi-Majorana modes of finite Majorana polarization appear at finite energy in the normal region. The extension of these modes in the normal region is smaller and smaller, their polarization goes to one and their energy goes to zero when the coupling between the normal and the SC region goes to zero.

\begin{figure}[ht]
\includegraphics[width=\columnwidth]{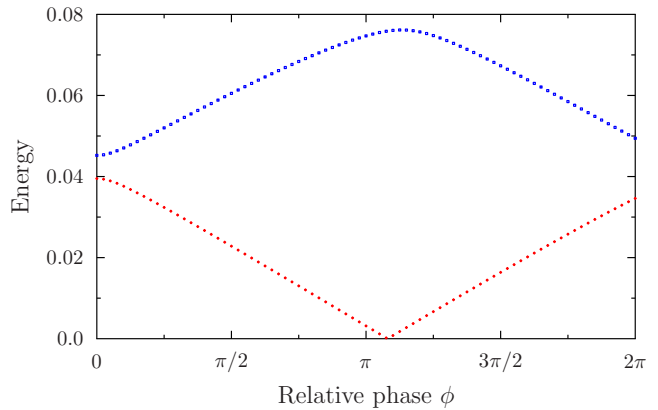}
\caption{(Color online) Evolution of the lowest-energy modes with the phase difference $\phi$. Note the shift of the $2\pi$ periodicity to a $2 \pi +2 \delta \phi$.}
\label{fig:grad}
\end{figure}

\begin{figure}[ht]
\includegraphics[width=\columnwidth]{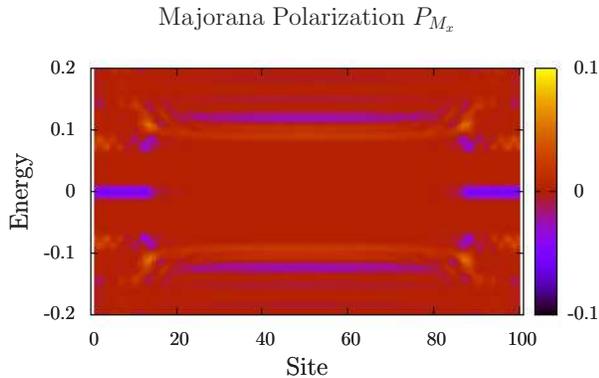}
\caption{(Color online) Majorana polarization as a function of energy and position for a ring with a total phase difference of $\pi+\delta\phi$. The system is periodic with the first and last site being equivalent. Extended Majorana modes form and are distributed over the entire normal region.}
\label{mpring}
\end{figure}

In what follows we will consider a perfect coupling between the normal and the SC state. While no Majorana states are formed when $\phi=0$, for peculiar values of the phase difference accumulated over the SC region, and for phase gradients that are not too large, Majorana fermions can form in the normal region. This can be seen in Fig.~\ref{fig:grad} where we plot the low-energy eigenvalues as a function of the total phase difference. We note that when this phase difference is slightly larger than $\pi$ one of the modes reaches zero-energy. 
We have checked that this is an actual Majorana mode by plotting the Majorana polarization as a function of position and energy across the ring (see Fig.~\ref{mpring}).  Same as in the previous analysis, the Majorana polarization of these extended modes is compensated by the Majorana polarization distributed continuously in the high-energy bands.

The difference $\delta\phi$ between $\pi$ and the phase corresponding to the formation of the Majorana mode stems from the presence of the gradient. This shift disappears for an infinite nanowire, corresponding to a vanishing gradient. We have checked explicitly that this "finite size effect" is reduced by increasing the size of the system.

This $2(\pi+\delta\phi)$ periodicity in the phase difference $\phi$ is preserved for gradients smaller than a critical value which for the chosen values of the parameters in our system is  $(\nabla\phi)_{gl}=0.27$, corresponding to a total phase difference of $\phi\simeq 21.6\pi$. As described above, larger gradients are predicted to drive the GR to a gapless phase. However our numerical simulations indicate that, while larger gradient values do seem indeed to take the system into the gapless phase, Majorana fermions may still survive for some peculiar gradient values; we do not understand what is the origin of this phenomenon.

\begin{figure}[ht]
\includegraphics[width=\columnwidth]{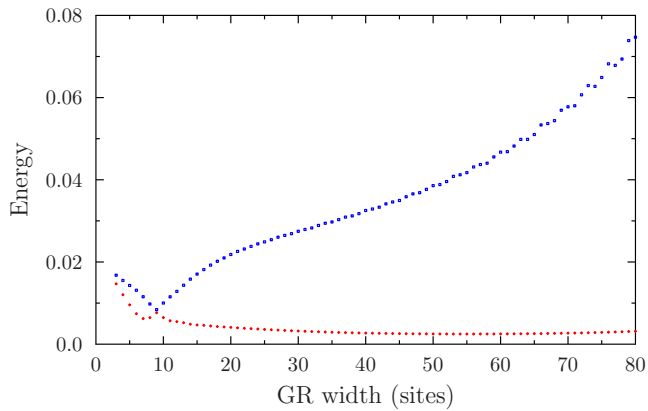}
\caption{(Colro online) The energy of the first two positive eigenvalues for a decreasing GR length. The total phase twist is kept to $\phi=\pi$, while the number of sites is varied. Close to the critical value of the gradient $(\nabla\phi)_{gl}=0.27 \approx \pi/11.6$, the system passes into a metallic regime. Due to finite-size effects our numerical analysis recovers this transition at a number of sites  of $9$ instead of $11-12$.}
\label{fig:dist}
\end{figure}

Another way to illustrate the evolution of the physics of the system with the value of the phase gradient is to fix the total phase gain $\phi=\pi$, and study the evolution of the low-energy modes with the number of sites in the GR (a larger number of sites is equivalent to a smaller phase gradient). As it has been illustrated in Fig.~\ref{fig:grad}, due to finite size effects, one has an infinitesimally-low energy mode; we note that the energy of this mode becomes smaller and smaller with increasing the number of sites. When the number of sites is reduced we expect to reach the critical value of the phase gradient that signals the passing of the system into the gapless phase. For an infinite system this is $(\nabla\phi)_{gl}\simeq 0.27$, corresponding to $\pi/11.6$. We would thus expect to see a phase transition when GR reaches the size of $11-12$ sites. Indeed a crossing of the bands and lifting of the low energy modes happens for a size of the GR of about $9$ sites (see Fig.~\ref{fig:dist}).

\section{Conclusion} \label{sec:conclusion}

We have used a tight-binding approach to calculate the energy spectrum, the local density of states and the Majorana polarization in topological SC wires in the presence of spin-orbit coupling and a Zeeman field. We have considered the effects of a non-zero penetration length of the SC in the normal state, as well as those of a uniform phase gradient in both linear and ring Josephson junctions. We have found that the ratio between the SC penetration length and the length of the wire can control the spatial extension of the Majorana modes in the normal section. 

We have also observed numerically the transition between the topological and gapless state for the section of the wire subject to a significant phase gradient, in both linear and ring junctions. For the linear junction, the only situation in which we have recovered extra Majorana states in the region subject to the phase gradient is when the phase gradient is chosen such that the phase difference between two adjacent sites is an odd multiple of $\pi$.  For the Josephson ring we have found that Majorana states can be induced in the normal region if the phase gradient is not too small, and if the total phase twist is close to an odd multiple of $\pi$. An interesting observation for systems subject to phase gradients is the formation of zero-energy Majorana modes whose full Majorana polarization is compensated by a continuous Majorana polarization distribution in the high-energy bands, such that the energy integral of the Majorana polarization on-site is equal to zero.  

\acknowledgments

The work of C.B. and D.C. was supported by the ERC Starting Independent Researcher Grant NANO-GRAPHENE 256965.

\end{document}